\documentclass[aps,preprint,showpacs]{revtex4}
\usepackage{amsfonts}
\usepackage{amsmath}
\usepackage{amssymb}
\usepackage{graphics}

\begin{document}

\preprint{MYaA/00134}
\title{Two-electron photoionization of endohedral atoms}
\author{M. Ya. Amusia}
\affiliation{Racah Institute of Physics, The Hebrew University, Jerusalem 91904, Israel;
A. F. Ioffe Physical-Technical Institute, St. Petersburg, 194021, Russia}
\author{E. Z. Liverts}
\affiliation{Racah Institute of Physics, The Hebrew University, Jerusalem 91904, Israel}
\author{V. B. Mandelzweig}
\affiliation{Racah Institute of Physics, The Hebrew University, Jerusalem 91904, Israel}

\begin{abstract}
Using $He@C_{60}$ as an example, we demonstrate that static potential of the
fullerene core essentially alters the
cross section of the two-electron ionization differential in one-electron energy $d\sigma ^{++}(\omega
)/d\varepsilon $. We found that at high photon energy prominent oscillations
appear in it due to reflection of the second, slow electron wave on the $%
C_{60}$ shell, which "dies out" at relatively high $\varepsilon $ values, of
about 2$\div $3 two-electron ionization potentials. The results were
presented for ratios $R_{C_{60}}(\omega ,\varepsilon)\equiv d\sigma ^{++}(\omega
,\varepsilon)/d\sigma ^{a++}(\omega,\varepsilon )$,
where $d\sigma ^{a++}(\omega,\varepsilon)/d\varepsilon$ is the two-electron differential photoionization cross section.
We have calculated the ratio $R_{i,ful}= \sigma_{i} ^{++}(\omega)/\sigma_{i}^{a++}(\omega)$, that accounts for reflection of both 
photoelectrons by the $C_{60}$ shell.
We have calculated also the value of two-electron photoionization cross section $\sigma ^{++}(\omega)$ and found that this value is close to that of an isolated $He$ atom.
\end{abstract}

\maketitle

\section{Introduction}

Elimination of two electrons by a single photon from an atom or multi-atomic
formation, such as cluster or fullerene, can take place only if the
inter-electron interaction is taken into account.

The desire to study the manifestation of this interaction stimulates
extensive experimental and theoretical investigation of the process that has
particularly intensified during last ten-fifteen years (see, e.g., \cite{1,
2}). Although a number of atoms were studied, the primary attention was
given to Helium. At this moment theoretical and experimental investigations
cover the frequency region from the near threshold region up to photon
energy $\omega $ much higher than the two-electron ionization potential $%
I^{++}$.

Very close to threshold the two-electron photoionization cross section is
determined by so-called Wannier regime \cite{3} with both electrons strongly
repulsing each other that results in acquiring almost the same energy and
moving in opposite directions.

With increase of $\omega $ for $\omega $ considerably more than $I^{++}$,
so-called shake-off (SO) mechanism \cite{4} became dominant, in which one
electron leaves the atom carrying away almost all energy $\omega $, while
the second is removed due to alteration of the field acting upon it after the
first electron emission. With energy growth so called quasi-free (QF)
mechanism \cite{5} becomes increasingly important. This mechanism accounts for
almost equal sharing of photon energy between photoelectrons, where their
interaction on the way out of the ionized atom is inessential.

Obviously, the cross section of two-electron photoionization $\sigma
^{++}(\omega )$ is determined by initial and final state wave function of
the considered object - target atom at the beginning and ion with two
continuous spectrum electrons at the end. The situation is simplified considerably at
high enough $\omega $, where cross sections and other characteristics of the
two-electron ionization are expressed via the initial state wave function
only.

In fact, one has to have in mind that two-electron ionization is a pure
three-body problem only for two-electron atom and ions. This process for any
other, more complicated objects is determined by more sophisticated wave
functions. This is why most of the attention is given usually to studies of
two-electron photoionization of $H^{-}$, $He$ and helium-like atoms. At high 
$\omega $ the $\sigma ^{++}$ and $\sigma ^{+}$ have similar $\omega $
dependence \cite{6}. Therefore it is convenient to characterize the process by
the ratio $R(\omega )=\sigma ^{++}(\omega )/\sigma ^{+}(\omega )$.

The SO leads for $\omega >>I^{++}$ to $R(\omega )=R_{SO}$ (see \cite{7} and
references therein). Inclusion of QF increases $R(\omega )$ considerably but
at high enough $\omega $ the ratio again reaches its $\omega $-independent
value, $R_{QF}>R_{SO}$ \cite{8}.

During last time a lot of attention is given to photoionization of not 
isolated atoms, but atoms encapsulated into the fullerene, mainly $C_{60}$
shell (see, e.g. \cite{9, 10}). The research in this area until now is pure
theoretical. But we are positive, that it will become an object of
experimental studies in not too distant future. It is well known, that $%
C_{60}$ radius $R$ is much bigger than that of the atoms staffed inside the
so-called endohedral atom, $A@C_{60}$.

The inclusion of the $C_{60}$ shell can affect the photoionization atom in
four directions. At first, and this is usually taken into account, the
photoelectron emitted by $A$ is reflected by the $C_{60}$ shell that "works"
as a static potential. At second, the $C_{60}$ shell has its internal degrees of
freedom \cite{11} that can be excited thus modifying the photoionization of the
atomic shells in $A@C_{60}$ in a similar way that one shell affects the
photoionization of the other in isolated atoms \cite{12}. Among the $C_{60}$
degrees of freedom the most prominent is the giant resonance that persists
not only in \ $C_{60}$ \cite{11}, but in its ions as well \cite{13}. Real or virtual
excitations of the giant resonance can modify the $A$ atom photoionization
cross section from almost complete screening at $\omega \rightarrow 0$ to
prominent enhancement, at $\omega $ in the vicinity of the giant resonance
maxima \cite{14, 15}. Note, that at high $\omega $ this screening becomes mainly
inessential.

Other two directions of $C_{60}$ influence upon the $A$ atom photoionization
is the direct knock out of $C_{60}$ electrons by the photoelectron from $A$
and the participation of $C_{60}$ electrons in the decay, both radiative and
non-radiative, of the vacancy, created in $A$ after photon absorption. Note
that this mechanism is similar to one of the frequently discussed channels
of the molecular vacancy decay. The corresponding two possibilities in
application to $C_{60}$ are not yet studied at all.

In this paper we consider a more complex process, namely two-electron
photoionization of $He@C_{60}$. We will concentrate on the high $\omega $
region. Of course, to investigate this process experimentally is more complicated than
one-electron photoionization. However, there are no indications whatsoever that such an
investigation is either impossible (or even extra-ordinary difficult) or
uninteresting.

The role of $C_{60}$ shell in two-electron photoionization is more complex
than in the single-electron, since outgoing electrons can be either both from the inner
atom $A$, or both from the $C_{60}$, or one from $A$ and the other from $C_{60}$.
However, by measuring two outgoing electrons in coincidence we can
distinguish all the different processes.

We will consider mainly high $\omega $. Having in mind that our aim is study the $%
C_{60}$ influence upon double ionization, we have to consider the SO
mechanism only. Indeed, the QF leads to two fast electrons, which are not
affected by $C_{60}$, while in SO one of the electrons is slow. Its
probability to leave $He@C_{60}$ can be therefore strongly affected.

Our attention in this paper will be given to differential in energy cross
sections $d\sigma^{++}(\omega)/d\varepsilon _s$, where $\varepsilon _s$ is
the "slow" photoelectrons energy. It is implied that the "fast" electron
energy $\varepsilon _f$ is given by the conservation law $\varepsilon
_f = \omega - \varepsilon _s - I^{++}$. Atomic system of units is accepted
in this paper: $e = m_{e} = \hbar = 1$, where $e$ and $m$ are electron
charge and mass, respectively.

We will assume that both electrons are removed from the $He$ atom, thus
intensionally neglecting a process that one can call "$A@C_{60}$ shake off",
in which after "fast" electron leaves $A$, the "slow" electron is emitted by
the $C_{60}$ shell instead of located inside atom $A$. The "$A@C_{60}$ shake off" is
potentially very important due to big number of available for removing
electrons. Therefore, it is possible that "$He@C_{60}$ shake off" is much
more probable than the ordinary one. It is possible, that an important role
can be played by multiple "$A@C_{60}$ shake of", in which not one but
several electrons can be removed from $C_{60}$ after instant creation of a
vacancy in $A$. One has to have in mind that for an isolated atom $A$ the
photoionization with excitation cross section $\sigma^{+*}(\omega)$ can be
of the order or even bigger than $\sigma^{++}(\omega)$. It is essential to
have in mind that the $\sigma^{+*}(\omega)$ for $A@C_{60}$ can be converted into $%
\sigma^{++}(\omega)$, since the excitation energy of $A^{+}$, particularly
in the case of $He$, is bigger than the ionization potential of $C_{60}$. We
plan, however, to concentrate on "$A@C_{60}$ shake off" in another paper.

\section{Main formulas}

The two-electron photoionization cross section of an atom $A$ in initial state $%
i $,$\ \sigma _{i}^{a++}(\omega ),$ can be presented by the following
expressions:

\begin{equation}\label{1}
\sigma _{i}^{a++}(\omega )=\int\nolimits_{0}^{\omega -I^{++}}\frac{d\sigma
_{i}^{a++}(\omega ,\varepsilon )}{d\varepsilon }d\varepsilon { ,}
\end{equation}

where at high $\omega $, $\omega >>I^{++},$ the differential in energy cross
section $d\sigma _{i}^{a++}(\omega ,\varepsilon )/d\varepsilon $ is given by
the expression \cite{16}:

\begin{equation}\label{2}
\frac{d\sigma _{i}^{a++}(\omega ,\varepsilon )}{d\varepsilon }=\frac{32\sqrt{%
2}Z^{2}\pi }{3c\omega ^{7/2}}|\int \Psi _{i}(0,\mathbf{r})\varphi
_{\varepsilon 0}(\mathbf{r})d\mathbf{r|}^{2}{.}
\end{equation}

Here $\Psi _{i}(0,\mathbf{r})$ is the $\mathbf{r}_{1}=0$ value of $\Psi _{i}(\mathbf{r}_{1},\mathbf{r}_{2}),$ which is in
our case the initial state wave function of atomic $He$. The "slow" outgoing
electron is described by a pure Coulomb wave function that describes an $s$
-wave electron that moves in the field of a nucleus with charge $Z$.

Similar to Eq.(\ref{1})  the cross-section of ionization with excitation $
\sigma^{+*}(\omega)$ is given by the following expression:

\begin{equation}\label{3}
\frac{d\sigma _{i}^{a++}(\omega ,\varepsilon )}{d\varepsilon }=\frac{32\sqrt{
2}Z^{2}\pi }{3c\omega ^{7/2}}\sum\limits_{n>0}{\vert\int \Psi _{i}(0,\mathbf{r}
)\varphi _{n0}(\mathbf{r})d\mathbf{r}\vert}^2 .
\end{equation}

Here $n$ is the principal quantum number of the second electron excitation
level.

Now let us consider $He@C_{60}$ two-electron photoionization and take into
account the $C_{60}$ shell. It is clear that $C_{60}$ potential does not
affect the "fast" electron, which remains to be described by a plane wave.
The $C_{60}$ shell generates potential field that is almost uniform at
atomic distances $r_{a}$, where the initial state of $He$ is located.
Therefore, for the initial state the embedding of an atom $A$ into $C_{60}$
can lead only to the shift of the energy scale, i.e., to modification of the
ionization threshold. This feature is inessential for us, since $\omega >>I^{++}$.

Thus, we have to take into account the action of $C_{60}$ upon the slow
outgoing electron. The fullerene is a very complex structure. Therefore, to
take into account its action upon photoelectron essential simplifications
are necessary. We will follow here the approach developed in a number of
papers (see \cite{17, 18} and references therein), that substitutes the
complicated $C_{60}$ multi-atomic structure by a very simple so-called
bubble potential $V(r)=V_{0}\delta (r-R)$, where $V_{0}$ is chosen to
reproduce the affinity energy of the negative ion $C_{60}$, and $\delta (r-R)$
is Dirac delta function.

As it was demonstrated in \cite{18} and \cite{19} in connection to the one-electron
photoionization, the influence of the bubble potential can be taken into
account analytically, by constructing the outgoing wave function as a
superposition of regular $\varphi _{\varepsilon l}(\mathbf{r})$ and
irregular $\chi _{\varepsilon l}(\mathbf{r})$ (singular at $\mathbf{r}%
\rightarrow 0$) solutions of the Schr\"{o}dinger equation for an electron
with energy $\varepsilon $. Inside the potential bubble the photoelectron
wave function $\psi _{\varepsilon l}(\mathbf{r})$\ differs from $\varphi
_{\varepsilon l}(\mathbf{r})$\ only by a normalization factor 
$D_{l}(\varepsilon ):\psi _{\varepsilon l}(\mathbf{r})=$\ $
D_{l}(\varepsilon )\varphi _{\varepsilon l}(\mathbf{r})$. 
The factor $D_{l}(\varepsilon)$ depends upon the 
photoelectron energy $\varepsilon $, with $l$ being its electron angular
momentum. Outside the $\delta $-sphere the function $\psi _{\varepsilon l}(%
\mathbf{r})$\ is a linear combination of $\varphi _{\varepsilon l}(\mathbf{r}%
)$ and $\chi _{\varepsilon l}(\mathbf{r})$.\ The coefficients of the linear
combination are defined by the matching conditions of the wave functions on
the spherical shell, i.e., at $r=R$.

In the $\delta $-potential approximation the differential two-electron
photoionization of $He@C_{60}$ is thus given by formula similar to (\ref{1}),
that includes, however the factor $D_{l}(\varepsilon )$:

\begin{equation}\label{4}
\frac{d\sigma _{i}^{++}(\omega ,\varepsilon )}{d\varepsilon }=\frac{32\sqrt{2%
}Z^{2}\pi }{3c\omega ^{7/2}}|D_{0}(\varepsilon )|^{2}|\int \Psi _{i}(0,%
\mathbf{r})\varphi _{\varepsilon 0}(\mathbf{r})d\mathbf{r|}^{2}{ .}
\end{equation}

The expression for $|D_{l}(\varepsilon )|^{2}$\ is derived in \cite{19, 20} and
is presented as

\begin{equation}\label{5}
|D_{l}(\varepsilon )|^{2}\equiv \frac{(k/\Delta L)^{2}}{%
[u_{kl}(R)v_{kl}(R)-k/\Delta L]^{2}+u_{kl}^{4}(R)}.
\end{equation}

Here $u_{kl}(R)$ and $v_{kl}(R)$ are functions connected to the radial parts
of regular $\varphi _{\varepsilon l}(\mathbf{r})$ and irregular $\chi
_{\varepsilon l}(\mathbf{r})$ functions by relations $u_{kl}(r)=r\varphi
_{\varepsilon l}(r)$ and $\nu _{kl}(r)=r\chi _{\varepsilon l}(r)$; $k=\sqrt{%
2\varepsilon }$, $\Delta L$ is the discontinuity of the logarithmic
derivative of the wave function at $r=R$, connected to the fullerene radius $%
R$ and the electron affinity $I_{f}$ of the empty $C_{60}$ through the
expression

\begin{equation}\label{6}
\Delta L=-\beta (1+\coth \beta R),
\end{equation}
where $\beta =\sqrt{2I_{f}}$. The formula obtained are valid for low enough
energies of the "slow" photoelectron. Namely, its wave length should be much
bigger than the thickness of the fullerene shell.

At first one should calculate the ratio of differential cross sections (\ref{4})
to (\ref{2}) $R_{C_{60}}(\varepsilon )$ that according to Eqs.(\ref{3}) and (\ref{4}) is
independent upon $\omega $

\begin{equation}\label{7}
R_{C_{60}}(\varepsilon )\equiv d\sigma _{i}^{++}(\omega ,\varepsilon
)/d\sigma _{i}^{a++}(\omega ,\varepsilon )=|D_{0}(\varepsilon )|^{2},
\end{equation}
and then turn to the ratio $R_{i,ful}$\ of cross sections $\sigma
^{++}(\omega )$ and $\sigma ^{a++}(\omega )$ that are determined using Eq.(\ref{1})
with $d\sigma _{i}^{++}(\omega ,\varepsilon )/d\varepsilon $ given by Eq.(\ref{4}) and
Eq.(\ref{2}), respectively.

\begin{equation}\label{8}
R_{i,ful}=\sigma _{i}^{++}(\omega )/\sigma _{i}^{a++}(\omega )
\end{equation}

It is evident from (\ref{7}) that determined in this way $R_{i,ful}$\ is $\omega $
-independent at high $\omega $.

Since $\varphi _{\varepsilon 0}(\mathbf{r})$ and $\chi _{\varepsilon l}(
\mathbf{r})$ are pure Coulomb functions, the functions $u_{kl}$ and $v_{kl}$
(from Eqs.(\ref{6},\ref{7})) are expressed via the regular $F_{l}(\eta ,\rho )$ and
irregular $G_{l}(\eta ,\rho )$ Coulomb wave function, respectively. Namely,
one has $u_{kl}(r)=F_{l}(-\frac{Z}{k},kr)~$and $v_{kl}(r)=G_{l}(-\frac{Z}{k}
,kr)$, i.e. regular and irregular Coulomb wave functions that can be found
in \cite{20}.

Let us note that the formula (\ref{7}) can be extended to lower $\omega $.
Indeed, the interaction between two ionized electrons is particularly
essential near ionization threshold $\omega \geq I^{++}$, where Wannier
regime prevails. It is known that already several $eV$ above threshold the
Wannier expressions are no more valid. Therefore, it is reasonable to assume that at
about $10eV$ the interaction between outgoing electrons is inessential. One
of them is represented by an $s$-wave, while the other by a $p$-wave, in
order to preserve the total initial state angular momentum that is zero for $%
He@C_{60}$ and $l=1$ for the photon. Due to action of the $C_{60}$ shell the
second electrons wave function is modified in the same way as $\varphi
_{\varepsilon 0}(\mathbf{r})$ in (2). Thus, it is reasonable to expect that
to take this action into account one has to introduce the factor $%
|D_{1}(\overline\varepsilon )|^{2}$ into (\ref{7}). As a result, one obtains instead
 of Eq.(\ref{7}) the following expression:

\begin{equation}\label{9}
R_{C_{60}}(\omega ,\varepsilon )\equiv d\sigma _{i}^{++}(\omega ,\varepsilon
)/d\sigma _{i}^{a++}(\omega ,\varepsilon )=|D_{0}(\varepsilon
)|^{2}|D_{1}(\omega -I^{++}-\varepsilon )|^{2}.
\end{equation}

Since for low and medium energies the cross section $d\sigma
_{i}^{a++}(\omega ,\varepsilon )/d\varepsilon $ cannot be calculated using
Eq.(\ref{2}), we have to use other sources of absolute values of it in order to
obtain $R_{C_{60}}(\omega )$.

As such, either experimental data or results
of existing calculations of differential in energy $\varepsilon $\ cross
section can be used to substitute into 

\begin{equation}\label{10}
R_{C_{60}}(\omega )\equiv \int\nolimits_{o}^{\omega
-I^{++}}|D_{0}(\varepsilon )|^{2}|D_{1}(\omega -I^{++}-\varepsilon
)|^{2}d\sigma _{i}^{a++}(\omega ,\varepsilon )/\int\nolimits_{o}^{\omega
-I^{++}}d\sigma _{i}^{a++}(\omega ,\varepsilon ).
\end{equation}

In order to have a crude estimation of the role of both factors in (\ref{9}), we
have calculated in Eq.(\ref{10}) the factor $R_{C_{60}}(\omega )$ using Eq.(\ref{1}) 
well outside its range of validity $\omega \gg I^{++}$, which gives us $R^{(h)}_{C_{60}}(\omega )$.

Since at  $\omega $ close to $I^{++}$, $d\sigma _{i}^{a++}(\omega ,\varepsilon )/d\varepsilon $
is almost $\omega $-independent, as an estimation for $R_{C_{60}}(\omega )$
at $\omega \geq I^{++}$ can serve an approximate relation

\begin{equation}\label{11}
R^{(l)}_{C_{60}}(\omega )\equiv \int\nolimits_{o}^{\omega
-I^{++}}|D_{0}(\varepsilon )|^{2}|D_{1}(\omega -I^{++}-\varepsilon
)|^{2}d\varepsilon /(\omega -I^{++}).
\end{equation}

\section{Details of calculations}

First of all, one should notice that the term $k/\Delta L$ in denominator
of Eq.(\ref{5}) was taken with the sign "minus", whereas there was "plus" in
the Ref.\cite{18,19}. This is because the authors of the latter references used
the irregular Coulomb functions $\overline {v}_{kl}(r)=-v_{kl}(r)$, whilst we used
the Coulomb functions $v_{kl}(r)$ presented in the Handbook \cite{20}.

The Eqs.(\ref{9}-\ref{11}) include the function $D_{1}(\overline\varepsilon)$ at $\overline\varepsilon=0$, for $\varepsilon=\omega -I^{++}$.
It is clear that the Coulomb functions $F_{l}(-\frac{Z}{k},kr)$ 
and $G_{l}(-\frac{Z}{k},kr)$, presented in the expression (\ref{5}) for $D_{l}^2$, 
have singularities at zero energy, i.e. at $k=0$ that corresponds to $\overline\varepsilon=0$.
However, the presentations of Coulomb functions given in \cite{20} enable to obtain
the following limit relations:

\begin{equation}\label{21}
\lim _{_{k\rightarrow 0}}\frac{u_{kl}(r)}{\sqrt{k}}\equiv f_{l}(Z,r)=\sqrt{\pi r}J_{2l+1}(\sqrt{8Zr}), 
\end{equation}
\begin{equation}\label{22}
\lim _{k\rightarrow 0}\frac{v_{k0}(r)}{\sqrt{k}}\equiv g_{0}(Z,r)=\frac{p(2Zr)}{
\sqrt{2\pi Z}}-\sqrt{\frac{r}{\pi }}\left [2\gamma -1+\ln (2Zr) \right ]J_{1}(\sqrt{8Zr}).
\end{equation}
Here \( \gamma  \) is Euler's constant, \( J_{\mu}(x) \) is a usual 
Bessel function of the first kind,
and \( p(\alpha ) \) is the following sum, which is infinite but rapidly convergent:
$p(\alpha )=\sum ^{\infty }_{j=0}P_{j}(\alpha )$, where $P_{j}(\alpha )$
could be obtained by the recurrence relations
\begin{equation}\label{23}
P_{j}=-\frac{\alpha }{j(j-1)}\left[ (-1)^{j}\frac{(2j-1)j\alpha ^{j-1}}{(j!)^{2}}+P_{j-1}\right]
\end{equation}
with the initial values $P_{0}=1,~~ P_{1}=0$. Functions $g_{l}(Z,r)$ for $l>0$ could be then obtained by using Eqs.(\ref{21})-(\ref{23}) and the following recurrence relation
\begin{equation}\label{24}
g_l(Z,r)f_{l-1}(Z,r)=g_{l-1}(Z,r)f_l(Z,r)+\frac{l}{Z}.
\end{equation}
The latter is a consequence of the Wronskian relation (see Eq.(14.2.5) in Ref.\cite{20}) for the regular and irregular Coulomb functions. Note that Eq.(\ref{21}) for the regular Coulomb functions could be found in numerous handbooks on non-relativistic quantum mechanics (see, e.g. Ref.(\ref{22})), within the accuracy of a normalization factor. However, we did not find in scientific literature the corresponding relations (\ref{22})-(\ref{24}) for irregular Coulomb functions.

Using the limit relations (\ref{21})-(\ref{24}), one could present Eq.(\ref{5}) for the zero energy in the form:
\begin{equation}\label{25}
D^{2}_{l}(0)=\frac{1}{[(\Delta L) g_{l}(Z,R)f_{l}(Z,R)-1]^{2}+(\Delta L)^{2}f_{l}^{4}(Z,R)}.
\end{equation}
In particular, for the helium atom expression (\ref{25}) yields: $D^{2}_{0}(0)=1.38590$ and $D^{2}_{1}(0)=3.09129$.

The function $\Psi _{i}$ can be obtained using variational principle that
minimizes the initial state energy.
One has to have in mind, that, generally speaking, this procedure is able to
reproduce initial state energy absolutely accurate, leading to wave function
that satisfies the precise Schrodinger equation for two electrons in the
field of a nucleus with the charge $Z$. In its practical implementations,
however, the variation wave functions reproduces the initial energy
approximately and therefore is good on the average.

As to photoionization, since one of the outgoing electrons is fast, its wave
function is a plane wave with a wavelength short in the atomic scale.
Therefore, photoeffect is able to test the short range behavior of $\Psi
_{i}(\mathbf{r}_{1},\mathbf{r}_{2})$, namely, as it is seen from Eq.(\ref{1}), $\Psi
_{i}(0,\mathbf{r})$.

 This is why instead of a variational, we use a locally
correct wave function \cite{22,23,24}, that describes with high accuracy the
two-electron photoionization \cite{16} and some average characteristics of the
initial state.

\section{Results of calculations}

Here we presented the results of our calculations for the two-electron photoionization of $He@C_{60}$. For this object the radius $R=6.64a.u.$ and the electron
affinity of the empty $C_{60}$ is equal to $I_{f}=2.65eV$. The initial state
of $He$ considered in this paper is $i=1s$, i.e. $l=0$. 
The results for $R_{C_{60}}(\varepsilon )$ from Eq.(\ref{7}) are presented in Fig.1
together with the function $D^{2}_{1}(\varepsilon)$ from the Eqs.(\ref{9}-\ref{11}).
As it was expected and similar to the case of one-electron photoionization, the ratio 
$R_{C_{60}}(\varepsilon )$ is a strongly oscillating function of $\varepsilon $.
\begin{figure}
\begin{center}
\resizebox*{0.8\textwidth}{0.4\textheight}{\includegraphics{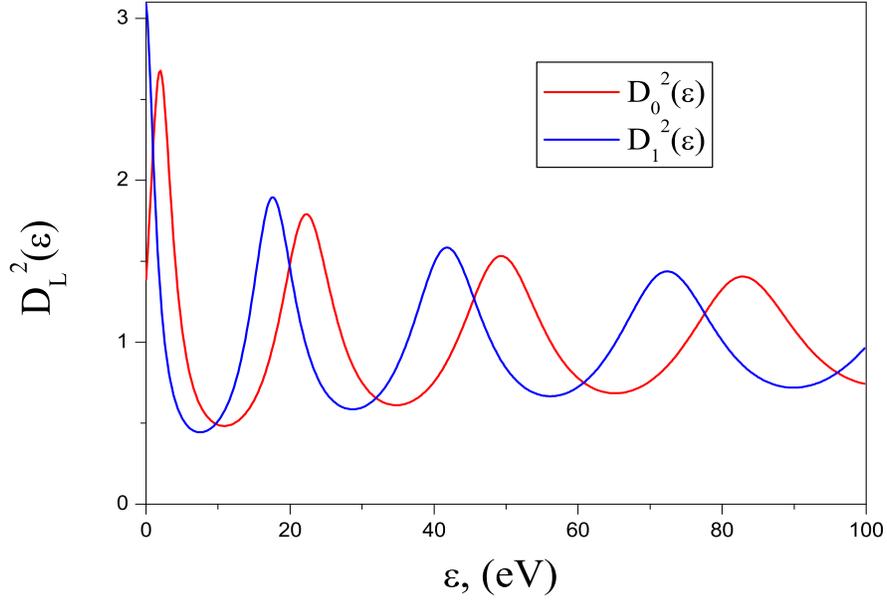}}
\end{center}
\caption{The reflection and refraction factors $D^{2}_{L}$ ($L=0,1$) as a function of photoelectron energy $\varepsilon$ for $He@C_{60}$.}
\end{figure}

Fig.(2) depicts $R_{C_{60}}(\omega,\varepsilon )$, given by Eq.(\ref{9}) for several $\omega$ as functions of $\varepsilon$. 
It is not incidental, that the curves behavior at $\varepsilon=0$ and 
$\varepsilon=\omega-I^{++}$ are different. This is because one of the electron
is represented by an $s$-wave while the other by $p$-wave. 
\begin{figure}
\begin{center}
\resizebox*{0.8\textwidth}{0.4\textheight}{\includegraphics{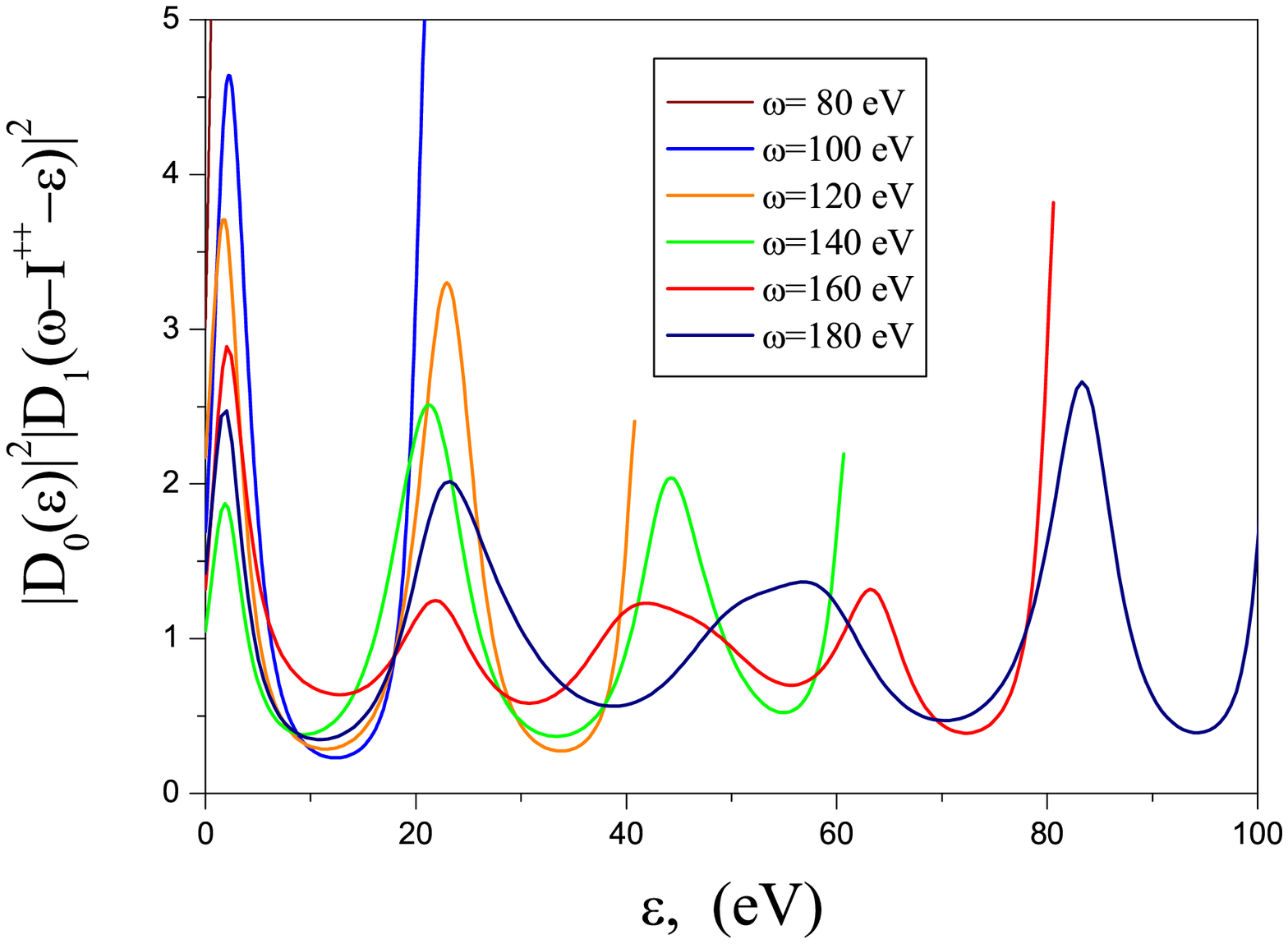}}
\end{center}
\caption{The cross sections ratio $R_{C_{60}}(\omega,\varepsilon)$ as a function of photoelectron energy
$\varepsilon$ for different values of photon energy $\omega$.}
\end{figure}

Note that $R_{C_{60}}(\omega,\varepsilon )$ becomes a prominently variating curve
already for $\omega-I^{++}\geq 20 eV$. For $\omega-I^{++}\geq 60 eV$ it has many
oscillations.
\begin{figure}
\begin{center}
\resizebox*{0.8\textwidth}{0.4\textheight}{\includegraphics{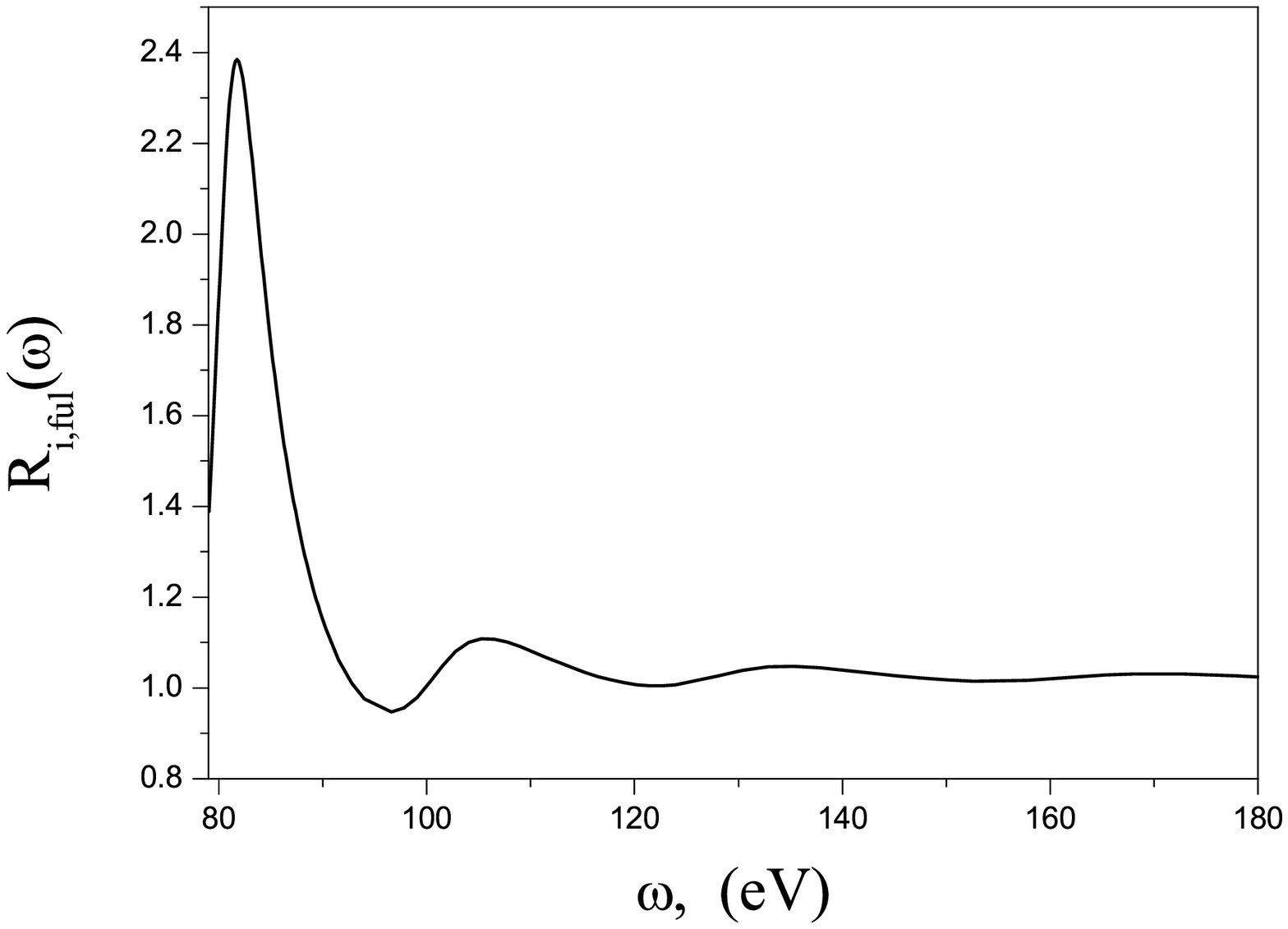}}
\end{center}
\caption{The cross sections ratio $R_{i,ful}$ as a function of photon energy
$\omega$.}
\end{figure}

Fig.(3) presents the results for the ratio $R_{i,ful}(\omega)$ determined by Eq.(\ref{8}). As one could see, with $\omega$ growth this ratio rapidly approaches
the asymptotic value $R_{i,ful}\rightarrow1$. Note, that in Eq.(\ref{8}) we assumed
that "slow" electron wave function is modified by the $C_{60}$ shell. In fact, 
this is incorrect already for $\omega\geq160 eV$.
\begin{figure}
\begin{center}
\resizebox*{0.8\textwidth}{0.4\textheight}{\includegraphics{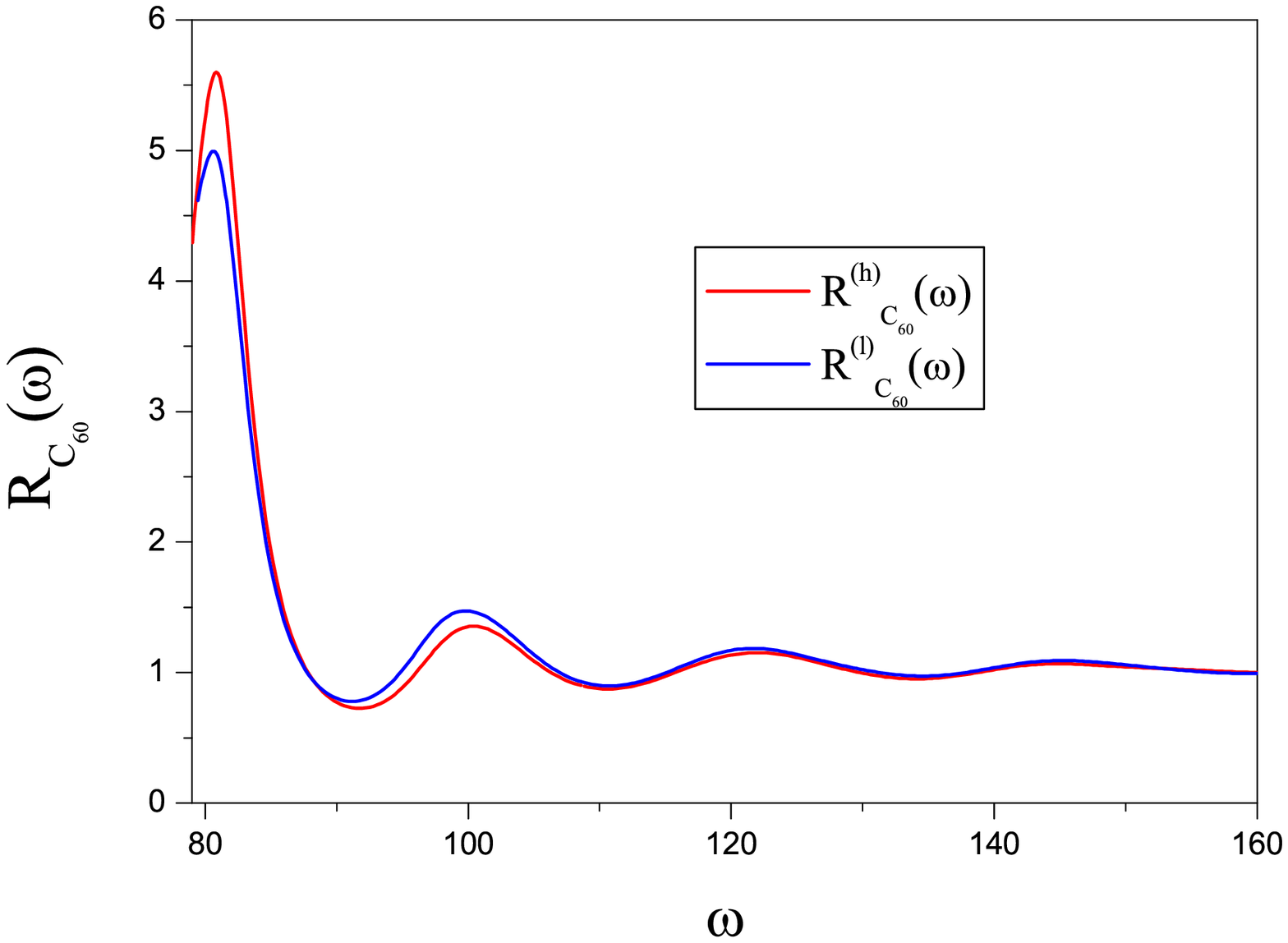}}
\end{center}
\caption{The cross sections ratios $R^{(h)}_{C_{60}}$ and $R^{(l)}_{C_{60}}$ as functions of  $\omega$ for high and low photon energy, respectively.}
\end{figure}

Fig.(4) presents the relations $R_{C_{60}}(\omega)$ given by Eqs.(\ref{10}-\ref{11}) for both
$\varepsilon$ distributions, that are valid respectively at high, and at low $\omega$.
It is seen that both curves, $R^{(h)}_{C_{60}}(\omega)$ and  $R^{(l)}_{C_{60}}(\omega)$,
prominently oscillate up to $\omega\sim150 eV$
and are close to each other. It permits us to suggest that the curve $R_{C_{60}}(\omega)$ is valid for correct $\varepsilon$ distribution instead of only
their limiting cases for high and low $\omega$.
The strongest is the role of the fullerene shell at $\omega< 110 eV$.

\newpage
 
\begin{acknowledgements}
M.Ya.A is grateful to Israeli Science Foundation, Grant No.174/03 for
support of this research. The research of V.B.M. was supported by
Grant No.2004106 from the United States-Israel Binational Science
Foundation (BSF), Jerusalem, Israel.
\end{acknowledgements}



\begin{thebibliography}{99}

\bibitem{1} J.~Ullrich, R.~Moshammer, R.~D\"{o}rner, O.~Jagutzki, V.~Mergel,
H.~Schmidt-B\"{o}cking and L.~Spielberger, J.\ Phys.\ B \textbf{30}, 2917
(1997).

\bibitem{2} H.\ McGuire, N.~Berrah, R.~J.\ Bartlett, J.~A.~R.\ Samson, J.~A.\
Tanis, C.~L.\ Cocke, and A.~S.\ Schlachter, J.\ Phys.\ B \textbf{28}, 913
(1995).

\bibitem{3} G.~H.\ Wannier, Phys.\ Rev. \textbf{90}, 817 (1953).

\bibitem{4} F.~W.\ Byron and C.~J.\ Joachain, Phys.\ Rev. \textbf{164}, 1
(1967).

\bibitem{5} M.~Ya.\ Amusia, E.~G.\ Drukarev, V.~G.\ Gorshkov, and M.~P.\
Kazachkov, J.\ Phys.\ B \textbf{8}, 1248 (1975).

\bibitem{6} M.~Ya.\ Amusia, \textit{Atomic Photoeffect}, Plenum Press, New York and
London (1990).

\bibitem{7} R.~Krivec, M.~Ya.\ Amusia, and V.~B.\ Mandelzweig, Phys.\ Rev.\ A 
\textbf{62}, 064701 (2000).

\bibitem{8} M.~Ya.\ Amusia, E.~G.\ Drukarev, and V.~B.\ Mandelzweig, Physica
Scripta (CAMOP), \textbf{72}, C22-29 (2005).

\bibitem{9} P. Decleva, G. De Ati, G. Fronzoni, and M. Stener, J. Phys. B.: \textbf{32},
4523 (1999).

\bibitem{10} M.~Ya.\ Amusia, A. S. Baltenkov, L. V. Chernysheva, Z. Felfli, and A.
Z. Msezane, J.\ Phys.\ B: At. Mol. Opt. Phys, \textbf{38}, L169-73 (2005).

\bibitem{11} I.~V.\ Hertel, H.~Steger, J. de\ Vries, B.\ Weisser, C.\ Menzel, B.\ Kamke
and W. Kamke, Phys.\ Rev.\ Lett. \textbf{68}, 784 (1992).

\bibitem{12} M.~Ya.\ Amusia, V.~K.\ Ivanov, N.~A.\ Cherepkov and L.~V.\ Chernysheva,
Phys.\ Lett.\ A, \textbf{40}, 5, p. 361 (1972).

\bibitem{13} S.~W.\ J.\ Scully, E.~D.\ Emmons, M.~F.\ Gharaibeh et. al., Phys.\ Rev.\ Lett., \textbf{94}, 065503, 1-4 (2005).

\bibitem{14} J.-P.\ Connerade and A.~V.\ Solov'yov, J.\ Phys.\ B \textbf{38}, 807 (2005).

\bibitem{15} M.~Ya.\ Amusia and A.~S.\ Baltenkov,
http://lanl.arxiv.org/abs/physics/0512269 (2005), Phys.\ Rev.\ A, submitted
(2006).

\bibitem{16} R.~Krivec, M.~Ya.\ Amusia, and V.~B.\ Mandelzweig, Phys.\ Rev.\ A, \textbf{63},052708 (2001).

\bibitem{17} A.~S.\ Baltenkov, J.\ Phys. B \textbf{32}, 2745 (1999).

\bibitem{18} M.~Ya.\ Amusia, A.~S.\ Baltenkov, and U.\ Becker, Phys.\ Rev.\ A \textbf{62}%
, 012701 (2000).

\bibitem{19} M.~Ya.\ Amusia, A.~S.\ Baltenkov, V.~K.\ Dolmatov, S.~T.\ Manson, and A.~Z.\ Msezane, Phys.\ Rev.\ A \textbf{70}, 023201-1-5 (2004).

\bibitem{20} M.\ Abramovitz and I.\ Stegun, \textit{Handbook of Mathematical Functions},
Dover, New York (1965). 

\bibitem{21} L.~D.\ Landau and E.~M.\ Lifshitz, \textit{Quantum mechanics : non-relativistic theory}, Pergamon Press, New York, 1991. 

\bibitem{22} M.~I.\ Haftel and V.~B.\ Mandelzweig,\ Ann. \ Phys.\ (N.Y.) 
\textbf{189}, 29 (1989). 

\bibitem{23} R.\ Krivec, V.~B.\ Mandelzweig, and K.~Varga, Phys.\ Rev.\ A \textbf{61
}, 062503 (2000).

\bibitem{24} E.~Z.\ Liverts, M.~Ya.\ Amusia, R.\ Krivec and V.~B.\ Mandelzweig, 
Phys.\ Rev.\ A \textbf{73}, 012514 (2006).

\end{thebibliography}
\end{document}